\documentclass{sig-alternate}
\usepackage[english]{babel}
\usepackage{times}
\usepackage{latexsym}
\usepackage{amsmath}
\usepackage{multirow}
\usepackage{graphicx}
\usepackage{url}

\begin{document}

\title{``Piaf'' vs ``Adele'': Classifying encyclopedic queries using
automatically labeled training data}

\numberofauthors{2} 
\author{
	\alignauthor Pedro Saleiro\\
       \affaddr{DEI-FEUP, University of Porto}\\
       \affaddr{Rua Dr. Roberto Frias, s/n}\\
       \affaddr{4200-465 Porto, Portugal}\\
       \email{pssc@fe.up.pt}
\and
	\alignauthor Lu\'{i}s Sarmento\\
 	   \affaddr{LIACC-FEUP, University of Porto}\\
       \affaddr{Rua Dr. Roberto Frias, s/n}\\
       \affaddr{4200-465 Porto, Portugal}\\
       \email{luis.sarmento@gmail.com}
}

\maketitle
\begin{abstract}
Encyclopedic queries express the intent of obtaining information typically available in encyclopedias, such as biographical, geographical or historical facts.
In this paper, we train a classifier for detecting the encyclopedic intent of web queries. For training such a classifier, we automatically label training data from raw query logs. We use click-through data to select positive examples of encyclopedic queries as those queries that mostly lead to Wikipedia articles. We investigated a large set of features that can be generated to describe the input query. These features include both term-specific patterns as well as query projections on knowledge bases items (e.g. Freebase). Results show that using these feature sets it is possible to achieve an $F_{1}$ score above 87\%, competing with a Google-based baseline, which uses a much wider set of signals to boost the ranking of Wikipedia for potential encyclopedic queries. The results also show that both query projections on Wikipedia article titles and Freebase entity match represent the most relevant groups of features. When the training set contains frequent positive examples (i.e rare queries are excluded) results tend to improve.
\end{abstract}

\category{}{Information Systems}{Query intent}

\terms{Experimentation}


\section{Introduction}

Web users have different search intents which they may or may not declare explicitly in the query string -- e.g. ``holidays in New York'' vs. ``booking hotel in Manhattan''. Inferring user intent in less explicit queries is crucial to improve several aspects of web search, such as direct results, ranking and suggestion. Broder \cite{Broder2002} represented user intent by classifying queries in three different categories: informational (e.g. ``anaemia symptoms''), navigational (e.g. ``nytimes'') and transactional (e.g. ``download free mp3''). We believe that it is possible and useful to further divide informational queries into more specific categories, namely \textit{encyclopedic queries}. 

We define \textit{encyclopedic queries} as those in which the user is aiming for information that can be typically found in a encyclopedia. These information needs often include biographical or geographical facts, information about historical events, well established scientific facts, etc. For instance, a user issuing the query ``\'{E}dith Piaf'' is probably aiming biographic information on Wikipedia about the singer personal life or career. On the other hand, a user issuing the query ``Adele'' is probably aiming Youtube videos of Adele performing ``Someone like you'' in the Brit Awards 2011 or the official Facebook page of the singer. The same comparison could be made with the queries ``1972 Olympics'' and ``2012 Olympics''. The former would be aiming information about the Munich massacre while the latter would be aiming news articles about the event or the 100 meters live stream.

Contrary to other subtypes of informational queries, e.g. questions or procedural queries, that focus on transient knowledge, \textit{encyclopedic queries} typically aim at more stable knowledge repositories. These repositories are becoming increasingly important in the Web, such as the Wikipedia. Furthermore, a significant number of queries leading to Wikipedia contain explicitly the word ``Wikipedia'' or abbreviations and misspellings of the word, e.g. ``wikipedia liz taylor'', ``kennedy weekpedia'' or ``cold war wiki''. Probably, users feel the need of boosting the Wikipedia position in the Search Engine Results Page (SERP) which is a sign that search engines are not completely well prepared to deal with non explicit \textit{encyclopedic queries}.

Therefore, identifying \textit{encyclopedic queries} is useful for tuning ranking functions or to decide when to present a direct result. However, this is a complex task due to the little query's context -- queries are often short (2-3 terms) and ambiguous -- as well as the broad number of domains related with encyclopedic knowledge. In fact, \textit{encyclopedic queries} are relatively frequent. A sample of query log data collected from the major Portuguese Web portal (sapo.pt) reveals that 0.89\% of clicks on SERP correspond to Wikipedia pages. Moreover, this percentage of clicks represents more than 3\% of distinct queries. For instance, queries about astronomy (``andromeda galaxy'') or second world war (``siege of leningrad'') lead to Wikipedia 100\% of the times. On the other hand, institutional or living celebrities queries, such as ``imf'' or ``tom hanks'', have several hits on many different web pages, exhibiting a click frequency of around 50\% on Wikipedia.

In this work, we present our first approach in training a classifier to automatically detect the encyclopedic intent of web queries. To train this classifier we automatically label training data from raw query logs, since manual annotation is expensive and it is not automatically adaptable to new trends. We propose to obtain positive and negative examples based on the Wikipedia click frequency of each query. Another challenge consists in selecting the most descriptive feature representation of the input queries for encyclopedic intent identification. Our first approach consists in combining both term-specific patterns with several fuzzy match conditions against knowledge bases, such as Freebase or Wikipedia article titles and category graph. Since the query log was provided by the biggest Portuguese Web portal (sapo.pt) the majority of the queries are written in Portuguese, Spanish and English. Therefore we used Wikipedia items written in all these languages.

The remainder of this paper is organized as follows: Section 2 introduces previous research work in web query intent, Section 3 presents our methodology followed by the experimental set-up in Section 4. Experimental results and analysis are described in Section 5. Finally, Section 6 presents the conclusions and future research directions.
 
\section{Related Work}

The user intent behind a search query has been subject of intense study during the last decade. Previous research works have represented user intent by query classification, by existing knowledge base (e.g. Wikipedia) or by queries and click-through data. The most common approach consists in representing user intent by classifying a query into a pre-defined category of goals/tasks or topics. Broder \cite{Broder2002} has proposed three categories representing the users' goal: navigational, transactional and informational queries. This has been the basis of several research works attempting to automatically identify the users' goal into these three categories \cite{Lee2005,Jansen2008}. Rose \cite{Rose2004} and Baeza-Yates \cite{yates2006} studied the users' intent problem and have proposed slight changes to the Broder's taxonomy. 

Li et al. \cite{Li2008} studied the problem of automatically obtaining large amounts of training data for query classification. The authors adopted a semi-supervised learning  approach in order to obtain large training sets using a bipartite graph representation of click-through data. In this way, the authors infer the class label of a given query by the proximity to labeled ones in the click graph. They applied this method to product intent and job intent classification obtaining F-measures of 74\% and 88\%, respectively. The winning solution of the 2005 edition of KDDCUP \cite{Shen2005} used search engine results as features and the Open Directory Project to create an intermediate taxonomy used for query topic classification into 67 pre-defined categories. The authors obtained an $F_{1}$ score of 42.3\%. 

More recently, Ji et al. \cite{Ji2011} defined search intent as a specific search task such as ``computer maintenance''. The authors identified a set of popular search tasks manually and then simultaneously classified queries and web pages into the search tasks. This classification used both query content and click-through data which were organized into a task-oriented graph. It is a two step approach, which consists in a topic classification followed by pre-defined task category classification. The authors applied the method to search tasks related to computers and cars achieving $F_{1}$ scores of 76\% and 79\%, respectively. 

Jethava et al. \cite{Jethava2011} presents an online algorithm for classifying web search intent described in terms of multiple dimensions (facets) such as topic, task or objective. They use a tree structured graphical model using only the query words. The WordNet database is also incorporated to identify previously unseen words. In the case of the facet task, authors classified queries as informational, non-informational and both, obtaining $F_{1}$ scores of 85\%, 72\%  and 15\% respectively. On the other hand, Hu et al. \cite{Hu2009} have used the Wikipedia as knowledge base. They mapped queries into articles and categories of Wikipedia which represent the user's intent. The authors obtained $F_{1}$ scores above 90\% both for travel, personal name and job intent classification.

Research works representing user's intent by queries and click-through data \cite{Cao2009,Shen2011,Ashkan2009} aim to exploit the user behavior as a source indicator of user's intent. Cao et al. \cite{Cao2009} create a conditional random field (CRF) model from the context information. The authors use neighboring queries in a search session and their clicked URLs as the representation of the user's intent. More recently Shen et al. \cite{Shen2011} proposed a sparse hidden-dynamics conditional random field (SHDCRF) model for learning user intent from search sessions. The authors classified user intent in 8 pre-defined categories achieving $F_{1}$ scores of 82\%. To the best of our knowledge no previous work has focused on the identification of the encyclopedic intent behind the search queries.

\section{Methodology}
The task we are addressing consists in building an \emph{Encyclopedic Intent Classifier} (EIC): given a query $q_{i}$ we want to classify it as an \emph{encyclopedic query} or not. We use a supervised learning approach to tackle this problem. In this section, we explain our approach by addressing three key aspects: (1) pre-processing of raw query logs, (2) automatic labeling of training data and (3) selecting the most appropriate feature representation for identifying encyclopedic intent. We consider only features extracted from query terms, without using any session data or user's browsing history. From now on we present our approach as the EIC.

\subsection{Query log pre-processing}
\label{subsec:preprocess}
Each line of an anonymized query log contains a tri-tuple  $<q_{i},h_{j},$ $ n_{k}>$, where $q_{i}$ is a query string (e.g.``academy awards 2011''), $h_{j}$ is the hostname of the clicked URL (e.g. ``oscars.org'') and $n_{k}$ is the click count (e.g. ``4'') of $<q_{i},h_{j}>$. In order to accurately process the query logs it is necessary to perform some pre-processing tasks. 

\subsubsection{Query string normalization}
Web users may submit queries in several different languages. However, regardless of the idiom of the query, we normalize the query terms by removing stop words and special characters, namely word accents and punctuation (quotation marks, brackets, etc.). Examples of non-normalized queries are ``beyonc\'{e} and jay-z'', ``The ballades of Chopin...'', ``vuelta+espa\~{n}a'' or ``latex/accents''. The only exception is question marks which are not removed (this is further detailed in Section 3.3.1).

\subsubsection{Navigational queries filtering}
Accordingly to Broder \cite{Broder2002} navigational queries depict the intent of visiting a particular web page (e.g ``facebook'' or ``nytimes''). We exclude navigational queries of our work because search engines are well prepared to deal with such queries and we are interested in distinguishing \textit{encyclopedic queries} within the group of informational queries. We exclude navigational queries using a simple heuristic. We consider a query as navigational if there is a term of the query string $q_{i}$ contained in the group of its clicked hostnames $H^{q_{i}}=\{h_{a},h_{b},...,h_{n}\}$. For example, the query ``amazon books'' is considered a navigational query because one of the clicked hostnames contains the term ``amazon'' namely, ``amazon.com''.

\subsubsection{Wiki queries filtering}
From now on, we refer to queries containing explicitly the word ``Wikipedia'' or abbreviations and misspellings of this word as \textit{Wiki} queries. For example, ``james dean wikpedia'', ``ancient greece wikipedia'' or ``wekpedia velvet revolution''. We believe that users feel the need of boosting the Wikipedia position in the Search Engine Results Page (SERP) however to avoid bias we exclude these queries from our experiment. The encyclopedic intent is explicitly expressed in these queries thus they would be easily classified as \textit{encyclopedic}. 

\subsection{Automatic labeling}
\label{subsec:autogen}
In this work we propose an annotation method based on click-through data mainly due to cost, i.e, manual annotation of millions of queries is expensive and adaptability, i.e., automatic annotation is able to adapt to time change conditions and user behavior evolution.

We annotate data by evaluating the click frequency on Wikipedia pages of each query. Therefore for each query $q_{i}$ we calculate the tuple $<C_{w}^{q_{i}},C_{\bar{w}}^{q_{i}}>$, where $C_{w}^{q_{i}} = \sum{n_{k}^{q_{i}}}$ if the hostname $h_{j}^{q_{i}} = ``wikipedia.org''$ and $C_{\bar{w}}^{q_{i}} = \sum{n_{k}^{q_{i}}}$ if the hostname $h_{j}^{q_{i}}\neq ``wikipedia.org''$. 
The ratio \[R_{w}^{q_{i}}=\frac{C_{w}^{q_{i}}}{C_{w}^{q_{i}}+C_{\bar{w}}^{q_{i}}}\]  is the click frequency on Wikipedia for a given query $q_{i}$. This ratio allow us to automatically label a query as \textit{encyclopedic}, $E$, or \textit{non-encyclopedic}, $\bar{E}$. Consequently, we define the \emph{encyclopedic} labeling criteria $C_{E}:Rw \geq \tau_{E}$ in which $\tau_{E}$ defines the lowest value of the ratio $R_{w}^{q_{i}}$ of \textit{encyclopedic queries}, $E$. For instance, ``palacio dos medicis'' (Medicis palace), ``europa de leste'' (Eastern Europe) or ``menuet'' represent examples of typical \textit{encyclopedic queries}, $E$. We define the \emph{non-encyclopedic} labelling criteria $C_{\bar{E}}:Rw \leq \tau_{\bar{E}}$ in which the threshold $\tau_{\bar{E}}$ ($\tau_{\bar{E}} < \tau_{E}$) defines the upper value of $R_{w}^{q_{i}}$ of \textit{non-encyclopedic queries}, \emph{$\bar{E}$}. Typical examples of \textit{non-encyclopedic queries}, \emph{$\bar{E}$} are ``resultados lotaria'' (lottery results), ``dietas'' (diets) or ``club amizade'' (friendship club).
Defining two evaluation criteria allow us to filter out queries which are not clearly classified as \emph{encyclopedic} or \emph{non-encyclopedic} based on their click frequency on Wikipedia pages. This method also allows including other repositories of information, as well as, different values for the decision thresholds.

\subsection{Query features}
We wish to explore the best combination of features to optimize encyclopedic intent classification. We investigate two type of features: (1) term-specific patterns which are obtained directly from the query string and (2) query projections in knowledge bases, namely web directories, Wikipedia page titles and Wikipedia category graph, as well as, the DBpedia ontology and Freebase categories.

\subsubsection{Term-specific patterns}
This type of features exploits several intrinsic characteristics of query terms both at semantic and morphological level. The approach consists in assigning a binary score if any of the following patterns occur in the query terms.
\begin{description}

\item[Dates in History:] \textit{Encyclopedic queries} often consist in queries about historical facts therefore we want to assess if there are date and time related terms, mainly months and years numbers in the query terms. Examples of queries exhibiting this pattern are ``tour france 1999'', ``guerra 1914'' (1914 war) and ``dia 10 de junho feriado'' (holiday June 10).

\item[Latin terms:] This pattern consists in the existence of Roman numerals and Latin suffixes in the query terms such as ``george III'', ``pintura dos seculos xi xiv'' (paintings from 14th 15th century), ``ficus aurea'' or ``otolemur crassicaudatus''. Roman numerals are common in the name of successive leaders (e.g. kings or popes) as well as in the name of annual events or references to past centuries. On the other hand, Latin suffixes (e.g. ``atus'',``arium'' or ``icus'') are usual in biological classification or medical terminology.

\item[Geographic terms:] Another common characteristic of \textit{encyclopedic queries} consists in queries about countries and cities information, such as a capital of a given country. We only considered cities and countries written in Portuguese, Spanish and English collected from Wikipedia ``List of countries'' and ``List of towns''. For example, ``capital of kazakhstan'', ``princess of Norway'' or ``spring in lisbon'' are queries containing geographic terms. 

\item[Query morphology:] We also explore some morphological patterns of the input queries such as the existence of question marks and the number of terms in a query. In some cases, \textit{encyclopedic queries} are in a form of a question (e.g. ``role play?'', ``protoestrela?'').  Regarding the query size, we define 6 different features corresponding to the number of terms in each query between 1 and 5 terms and more than 5.
\end{description}

\subsubsection{Knowledge base projections}

This type of features take advantage of knowledge bases content and structure in order to represent an input query. 

\begin{description}
\item[Web directories:] We exploit semantic lists from web directories, namely Wiktionary\footnote{http://pt.wiktionary.org} and Sapo Listas\footnote{http://services.sapo.pt/InformationRetrieval/SemanticLists} which consists in a Portuguese lexicon of 3128 words and 49 categories. We dynamically generate features representing the semantic category of each query term. For each query $q_{i}$ we submit its query terms to the semantic lists which, in the case of an exact match, retrieve a list of semantic categories, such as job category, type of organization or nationality. These categories represent semantic features of the query and typically grow in number with the dataset size.

\item[Wikipedia titles:] Articles are the atomic elements of Wikipedia. Each one describes a specific topic and its title is a string summarizing the topic under description. We extract this type of features by calculating the projection of a query in Wikipedia page titles. We use 9 different lists of page titles, corresponding to titles of articles (e.g. ``Commonwealth of Nations'' or ``Cretaceous''), titles of disambiguation pages (e.g. ``Primate'' or ``Mice'') and category strings (e.g. ``Historical continents'' or ``Electronics companies'') from Portuguese, Spanish and English Wikipedia pages. For each query $q_{i}$ we calculate its similarity with a Wikipedia page title string $w_{j}$ by calculating the Dice's coefficient: \[s = \frac{2*|q_{i} \cap w_{j}|}{|q_{i}|+|w_{j}|}\] 
For each page type and language we select the highest value of $s$ for $q_{i}$ and assign it to one of six different value intervals from 0 to 1. Therefore, we extract a feature based on the type of page, language and value of $s$. In summary, at most this type of features correspond to 54 distinct features (3 types of pages, 3 languages and 6 $s$ value intervals). For instance, the query ``viking age'' has an exact match ($s=1.0$) with a Wikipedia article title written in English. Therefore it would have the feature \emph{title-EN-1.0}.

\item[Wikipedia category graph:] Each Wikipedia article belongs to one or more categories which in turn are organized in a hierarchical semantic structure. We create a graph $G=(T,C)$ of the Wikipedia articles and categories network, such as Hu \cite{Hu2009}. We consider each node as a semantic class and consequently as a semantic feature. Thus, we extract high-level semantic features from the query projection in Wikipedia page titles. For each query $q_{i}$ we calculate the intersection scores $s$ for each Wikipedia page title of Wikipedia and choose the Wikipedia title $w_{j}$ corresponding to the highest score $s$ if $s > 0$.  Due to the Wikipedia category graph dimension and redundant connections we opt to extract as features, all the categories strings only in 4 levels of depth in $G$ for the given $w_{j}$. For instance, the query ``ideia deus taoismo'' (concept of god in Taoism) corresponds to an $s = 0.25$ for the Wikipedia title ``Taoism''. We extract all the category strings in 4 levels of depth in $G$ for the given $w_{j}$ . If we consider the previous example of ``Taoism'' we would extract 295 semantic classes, i.e., category strings from 4 levels of depth in $G$ from ``Taoism''. Examples of such categories are ``Classical Chinese philosophy'', ``Classical pantheism'' or ``Religious faiths, traditions, and movements''.

\item[DBpedia ontology:] DBpedia\footnote{http://dbpedia.org} created an ontology of Wikipedia articles where each title correspond to a semantic category such as ``Person'' or ``Place''. We take advantage of this resource and extract as features the semantic categories of a Wikipedia article which has an intersection score $s >0$ between its title $w_{i}$ and a given query. 

\item[Freebase entities:] Freebase\footnote{http://www.freebase.com} is an open repository of almost 23 million entities (persons, places or things). It allows to query the entity repository through a free text search API which returns the ranked list of similar entities in case of a match or an empty string if there is no entity match. We extract a binary feature consisting in success or failure of query match in Freebase entity repository. Additionally, we extract the top category of the result. For instance the query ''depeche mode band`` returns an entity with top category ''/music/`` encapsulated in the id field of query response.

\end{description}

\section{Experimental Set-up}
In this section, we describe the query log used in this work, the datasets obtained, as well as the evaluation metrics and baseline used to validate the effectiveness of our \textit{encyclopedic intent classifier}, EIC.
\subsection{Query log}
The query log is a sample of a real query stream collected between November 1, 2010 and January 16, 2012 from one of the main search portals in Portugal (www.sapo.pt). The log corresponds to more than 8 million distinct query strings which generated more than 53 million clicks on SERP. Most of the queries were written in Portuguese. This log includes navigational (e.g. ``hotmail''), transactional (e.g. ``download sp3 windows xp'') and informational queries (e.g. ``fine art definition''). For instance, navigational queries addressing Facebook represent 4.33\% of total clicks on SERP, counting 2,318,046 clicks. On the other hand, informational queries leading to Wikipedia comprise 477,830 clicks, representing 0.89\% of total clicks. Table 1 summarizes some relevant parameters of the query log.

\begin{table}[h]
\centering
\caption{Query log statistics.}
\renewcommand{\arraystretch}{1.2}
\begin{tabular}{l l}\hline
I. Query strings & 8,885,370\\
II. Total clicks & 53,539,801\\
III. Wikipedia clicks & 477,830 (0.89\% of II)\\
IV. Normalized queries & 6,857,480\\\hline
\end{tabular} 
\end{table}

The normalization process described in Subsection~\ref{subsec:preprocess} resulted in 6,857,480 distinct normalized queries. By calculating the ratio $R_{w}^{q_{i}}$ (Subsection~\ref{subsec:autogen}) we verify that 1.041\% of distinct queries always lead to Wikipedia, such as {}``blindness novel by Jos\'{e} Saramago'', {}``e=mc2'' or {}``1972 Olympics''. On the other hand, 96.81\% of distinct queries exhibit a ratio $R_{w}^{q_{i}}$ equal to 0, i.e, do not generate any click on Wikipedia. Examples of such queries are {}``playing chess online'', {}``viva la vida coldplay download'' or {}``f1 online streaming''. Another group of queries is composed by those having a ratio $R_{w}^{q_{i}}$ around 50\%. These queries usually lead to clicks in multiple sources and types of information (images, videos, news, etc.). Living celebrities, TV shows, corporations or famous products such as {}``gordon ramsey'', {}``monty python'' or {}``mercedes w210'' are some examples. In Figure 1, we show the query distribution based on the ratio $R_{w}^{q_{i}}$. 

\begin{figure}[h]
\includegraphics[width=1.0\columnwidth]{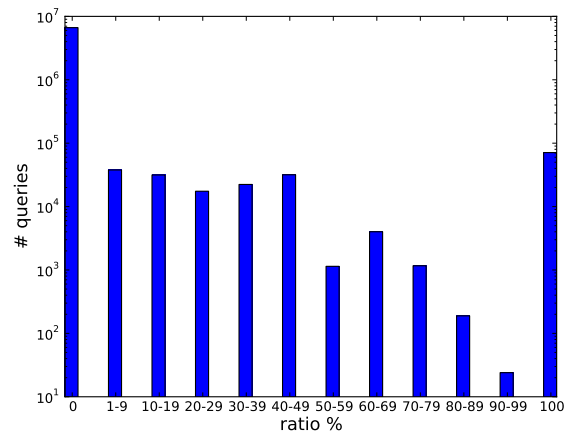}
\caption{$R_{w}^{q_{i}}$ distribution.}
\end{figure}
\subsection{Datasets}

We applied the method described in Subsection~\ref{subsec:autogen} to automatically annotate data. We defined a threshold $\tau_{E}=1.0$ for \textit{encyclopedic queries} as well as a threshold $\tau_{\bar{E}}=0.0$ for \textit{non-encyclopedic queries}. Let us call the set of \textit{encyclopedic queries} $Q_{E}$ while the set of  \textit{non-encyclopedic queries} will be called $Q_{\bar{E}}$. The set of \textit{encyclopedic queries} $Q_{E}$ contains 71,120 queries as depicted in Table 2. Examples of queries in $Q_{E}$ are  ``orthoptera'', ``capital of Thailand'' or ``biography of Yul Brynner''. On the other hand, the set of \emph{non-encyclopedic queries} $Q_{\bar{E}}$ comprises 6,638,964 queries, representing 96.81\% of total normalized queries. For instance, ``michelin tires price'', ``low cost flights to porto'' or ``restaurant fu jeng'' are examples of queries in $Q_{\bar{E}}$.

We filtered out 9,333 \emph{Wiki queries} from $Q_{E}$, i.e., queries explicitly targeting Wikipedia (e.g. ``wikpedia don johnson'' or ``madrid wiki'') as described in Subsection 3.1.3. In the same way, we removed navigational queries (e.g. ``mtv portugal'' or ``publico.pt'') from $Q_{\bar{E}}$. We detected 2,124,156 navigational queries in $Q_{\bar{E}}$, comprising 31.99\% of total queries $Q_{\bar{E}}$, as depicted in Table 2.

\begin{table}[h]
\centering
\caption{$Q_{E}$ and $Q_{\bar{E}}$ statistics.}
\renewcommand{\arraystretch}{1.2}
\begin{tabular}{l l}\hline
I. $Q_{E}$ queries & 71,120 \\ 
II. $Q_{\bar{E}}$ queries & 6,638,964 \\
III. \emph{Wiki queries} & 9,333 (13.12\% of I)\\
IV. Navig. queries & 2,124,156 (31.99\% of II)\\\hline
\end{tabular} 
\end{table}

\begin{figure}[h]
\includegraphics[width=1.0\columnwidth]{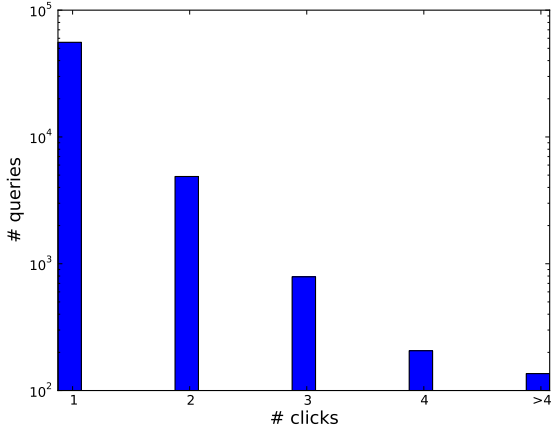}
\caption{Click distribution of $Q_{E}$}
\end{figure}

Figure 2 illustrates click distribution of $Q_{E}$ after removing \emph{Wiki queries}. The vast majority (55,780) of queries in $Q_{E}$ have just one click on Wikipedia pages. We not considered queries with only 1 or 2 clicks on Wikipedia pages because they represent the intent of a single user or at most, two users. Consequently, we created three different datasets: I, II and III.  For each dataset, we selected positive examples from $Q_{E}$ with at least 3, 4 and 5 clicks on Wikipedia pages respectively for Dataset I, II and III, as depicted in Table 3. We balanced the datasets by randomly selecting negative examples from $Q_{\bar{E}}$ in the same number as positive examples.

\begin{table}[h]
\centering
\caption{Datasets parameters.}
\renewcommand{\arraystretch}{1.2}
\begin{tabular}{c c c}\hline
Dataset & Size & Min. click freq.\\\hline
I & 2,262 & 3\\
II & 684 & 4\\
III & 272 & 5\\\hline
\end{tabular} 
\end{table}

\subsection{Wikipedia data}
The Wikimedia project provides access to periodically updated Wikipedia database dumps\footnote{http://dumps.wikimedia.org/}. As described in Section 3.3.1, we use Wikipedia articles titles, categories and disambiguation pages, in Portuguese, English and Spanish. The versions used in our experiments were released in January 2012. For sake of integrity the Wikipedia dumps were submitted to the same normalization process as the query log (Subsection 3.1.1). Table 4 summarizes the most relevant statistics of the database dumps used in our experiments.

\begin{table}[h]
\caption{Wikipedia data statistics.}
\centering
\renewcommand{\arraystretch}{1.2}
\begin{tabular}{ l  c  c  c }\hline 
Item & Portuguese & English & Spanish \\\hline
Articles  & 1,272,941 & 9,057,734 & 2,163,900\\
Disambig.  & 14,443 & 550,695 & 120,688\\
Categories  & 204,670 & 739,620 & 328,836\\ \hline
\end{tabular}
\end{table}

\subsection{Evaluation metrics}
We use classical evaluation metrics Accuracy, Precision, Recall and $F_{1}$ score to evaluate the effectiveness of an encyclopedic intent classifier. We define the evaluation metrics as follows:
\[
\textbf{Accuracy} = 
\]
\[
\frac{\# \textit{correctly classified }E + \# \textit{correctly classified }\bar{E} }{\#\textit{ total queries}}
\]
\[
\textbf{Precision} = \frac{\#\textit{correctly classified queries} E}{\# \textit{queries classified as } E}
\]
\[
\textbf{Recall} = \frac{\# \textit{correctly classified queries } E}{\#\textit{total queries } E}
\]
\[
F_{1}\textbf{ score} = 2 * \frac{Precision * Recall}{Precision + Recall}
\]

\subsection{Baseline}

We validate the \emph{Encyclopedic Intent Classifier} (EIC) by performing an empirical comparison with a Google-based baseline for all datasets. We submit each query $q_{i}$ on a given dataset to Google using the Google Custom Search REST API. Usually, 10 ranked URLS are retrieved. We classify $q_{i}$ as \textit{encyclopedic} if the hostname of the first result is ``wikipedia.org''. If this condition is not satisfied, $q_{i}$ is classified as \textit{non-encyclopedic}. In this way, we compute the standard performance metrics (Accuracy, Precision, Recall and F1) for a Google-based \textit{encyclopedic intent classifier}. We name this baseline method as BASE.

\section{Results and Analysis}

We apply two different classification algorithms: Support Vector Machines (SVM) \cite{svm} and Random Forests (RF) \cite{Breiman}. Both algorithms are effective in high dimensional spaces while memory efficient. We use a linear kernel function of SVM with the penalty parameter \textit{c} = 1.0 of the error term. In the case of RF we use 20 trees. For each dataset and classification algorithm, we perform a 10 K-fold cross validation. The results reported in Subsection 5.1 are the average of total runs.

\begin{table*}[!ht]
\centering
\caption{Comparison of datasets features.}
\renewcommand{\arraystretch}{1.2}
\begin{tabular}{ c  c  c  c  c}\hline 
Dataset & Total Features & DbPedia & Wikipedia Graph & Freebase\\ \hline
I & 25143 & 291 & 24592 & 118\\
II & 11408 & 234 & 10972 & 59\\
III & 6216 & 183 & 5717 & 35\\\hline
\end{tabular}
\end{table*}

\begin{table*}[!ht]
\centering
\caption{Comparison of EIC experimental results and BASE.}
\renewcommand{\arraystretch}{1.2}
\begin{tabular}{ cc c  c  c  c  c }\hline 
Dataset & Classifier & Accuracy & Precision & Recall & $F_{1}$ score\\ \hline
\multirow{3}{*}{I} 
& BASE & \textbf{82.45}\% & \textbf{90.87}\% & 72.76\% & 80.81\%\\ 
& EIC-SVM & 77.37\% & 75.29\% & \textbf{81.61}\% & 78.29\%\\
& EIC-RF & 79.49\% & 74.61\% & \textbf{89.57}\% & \textbf{81.39}\%\\ \hline

\multirow{3}{*}{II} 
& BASE & \textbf{82.75}\% & \textbf{91.73}\% & 72.86\% & 81.21\%\\ 
& EIC-SVM & 80.98\% & 78.29\% & \textbf{86.58}\% & \textbf{82.02}\%\\
& EIC-RF & 81.87\% & 77.38\% & \textbf{90.39}\% & \textbf{83.26}\%\\\hline

\multirow{3}{*}{III} 
& BASE & 83.03\% & \textbf{91.23}\% & 74.29\% & 81.89\%\\ 
& EIC-SVM & \textbf{84.56}\% & 82.58\% & \textbf{88.90}\%  & \textbf{85.18}\%\\
& EIC-RF & \textbf{86.79}\% & 84.54\% & \textbf{91.10}\%  & \textbf{87.27}\%\\\hline

\end{tabular}
\end{table*}

\subsection{Results}

Table 5 presents the number of features extracted during our experiments for all datasets. As expected, the total number of distinct features increases with the dataset size. Features obtained from Wikipedia graph represent more than 90\% of total number of distinct features in all datasets. On the other hand, features corresponding to categories of DBpedia ontology and Freebase top categories represent a less variable number of features due to the smaller number of categories when comparing with Wikipedia.

In Table 6, we report the experimental results obtained using Support Vector Machines (EIC-SVM), Random Forests (EIC-RF) and the Google-based baseline (BASE) for the three datasets. The results show that EIC improves with the click frequency on Wikipedia of the positive examples. The $F_{1}$ score for dataset I is 78.29\% (EIC-SVM) and 81.39\% (EIC-RF) while it is 85.18\% (EIC-SVM) and 87.27\% (EIC-RF) for dataset III. Dataset I contains positive examples with at least 3 clicks on Wikipedia while dataset III has a more restrict minimum click frequency -- 5 clicks. BASE exhibits constant performance regardless of click frequency on Wikipedia of the positive examples. This happens because Google was collecting, for a long period, statistics based on millions of queries on each day and thus almost every query is frequent enough in Google query logs. 

EIC exhibits better Recall results for all datasets while BASE has better Precision. In terms of Accuracy, BASE presents better results for dataset I and II while EIC-SVM and EIC-RF have better Accuracy in dataset III. In genneral EIC using Random Forests has better results than EIC using Support Vector Machines. The only exception is Precision in datasets I and II.

EIC presents $F_{1}$ scores above BASE for all datasets and using both classification algorithms, except in the case of EIC-SVM for dataset I. This Google-based baseline is extremely competitive, mainly because Google uses a wide set of signals besides any feature extracted from the query string in opposition to what EIC does. However, the major goal of using a Google-based baseline is to show that it is possible to achieve competitive results in the context of encyclopedic queries classification using a supervised learning approach with features extracted from the query string. Therefore, it would be useful for search engines to classify the query string and consequently, adapt the search interface, enhancing and complementing heavy processing tasks such as, consulting specific search indexes.

\subsection{Error analysis}
In this subsection, we detail the type of errors EIC performed on dataset III. The majority of false positives directly match with a Wikipedia title, however users have opted to select different repositories of information on the search engine results page (SERP). In fact for these queries, Wikipedia appears only in 4rd or 5th position on SERP. We identified some of the selected repositories, namely IMDB, Last.fm, Yahoo Answers, repositories of sports and medical information, Youtube videos and tutorials. Example of queries leading to these type of repositories are ``four weddings'', ``gregorians'', ``octreoscan'', ``sorteio quartos de final ta\c{c}a de portugal'' (Portuguese cup quarter finals draw) or ``1 metro s\~{a}o quantos cent\'{i}metros?'' (how many centimeters correspond to one meter?). In fact, most of false positives are \textit{encyclopedic queries} but we defined a strict labeling criteria based on click frequency on Wikipedia pages. In future works we shall consider using the click frequency on more online repositories of information. 

On the other hand, false negatives consist mainly in queries which contain Wikipedia concepts combined with another term(s). However, we would reduce recall if we tried to relax the matching criteria of Wikipedia projection of query strings. Therefore it is necessary to train our classifier with more frequent examples, as well as, with a larger number of examples. Another source of false positives consists in spelling errors, such as ``bety poob'', ``jormugand'', ``marie currie'' or ``churchil''. In this case, it would be useful to apply spelling correction methods. We have performed some experiments consisting in correcting any spelling error of dataset III which allowed us to obtain an average increase of 2\% in $F_{1}$ score. 

\begin{table*}[!ht]
\centering
\caption{Impact of removing each group of features (EIC-RF in Dataset III).}
\renewcommand{\arraystretch}{1.2}
\begin{tabular}{ c  c  c  c  c  c}\hline
\multirow{2}{*}{Features} & Affected & Accuracy & Precision & Recall & $F_{1}$ score\\
& queries & (\% diff)& (\% diff)& (\% diff)& (\% diff)\\ \hline
(-)Web directories & 105 & -1.43 & -1.90 & -0.83 & -1.18\\
(-)Term patterns & 272 & -1.82 & \textbf{-2.60} & -0.72 & -1.47\\
(-)Wikipedia Articles & 252 & \textbf{-3.66} & \textbf{-4.75} & \textbf{-2.14} & \textbf{-3.37}\\
(-)Wikipedia Disambig. & 227 & -1.45 & \textbf{-2.68} & +0.77 & -0.95\\
(-)Wikipedia Categories & 241 & -0.71 & \textbf{-2.10} & +1.54 & -0.22\\
(-)Wikipedia graph & 245 & \textbf{-2.17} & -0.76 & \textbf{-4.51} & \textbf{-2.37}\\
(-)DBpedia ontology & 244 & -1.82 & \textbf{-2.32} & -1.43 & -1.61\\
(-)Freebase & 111 & \textbf{-2.93} & \textbf{-3.49} & \textbf{-2.20} & \textbf{-2.70}\\\hline

\end{tabular}
\end{table*}

\subsection{Feature impact analysis}
In this subsection, we discuss the impact of different group of features in overall performance of our \emph{encyclopedic intent classifier} -- EIC, using Random Forests (EIC-RF) on dataset III. Table 7 includes the number of queries affected by removing a given group of features and the difference in the average results of Accuracy, Precision, Recall and $F_{1}$ score compared with the overall features results. 

The group of features affecting less queries is the \textit{Web directories} while the group of features \textit{Term Patterns} (includes \textit{Dates in History}, \textit{Latin terms}, \textit{Geographic terms} and \textit{Query morphology}) affects all queries. This happens because a feature related with query size is activated for every query. The results show that removing any group has a negative impact in $F_{1}$ score. This means that all group of features have valuable information  about the encyclopedic intent of the queries. 

Features based on the projection of queries in Wikipedia article titles (\textit{Articles titles}) represent the most significant group of features, exhibiting a negative impact of 4.75\% in Precision which leads to an impact on $F_{1}$ score of 3.37\%. The second most relevant group of features is the features extracted from Freebase which reveals the second most significant impact on all evaluation metrics, corresponding to 2.93\% on Accuracy and 2.70\% on $F_{1}$ score. 

The two most significant group of features in terms of Precision are Wikipedia article titles and Freebase while Wikipedia graph is the most significant group of features regarding Recall. Removing features representing projections on titles of disambiguation pages and category names of Wikipedia has a positive impact on EIC Recall but it still represents a negative impact on $F_{1}$ score.

\section{Conclusions and Future Work}
 Inferring user intent behind a search query is important to optimize several aspects of web search. Data collected from a real search engine (more than 6 millions queries) shows that more than 3\% of distinct queries lead at least once to a Wikipedia article and there are a significant number of users that explicitly request Wikipedia results. We have identified \textit{encyclopedic queries} among informational queries. \textit{Encyclopedic queries} represent the need for obtaining information usually available in more stable repositories, such as biographies, scientific facts or geographical information.
 
 We present our first approach towards an encyclopedic intent classifier. We leverage on automatically labeled training data which is cheaper and automatically adaptive to time change conditions when comparing with manual annotation of millions of queries. We find that a relatively simple encyclopedic intent classifier using only features based on query terms is able to compete with a Google-based baseline which uses millions of signals besides the query terms. Moreover, we believe there are significant benefits in applying supervised learning approaches using features extracted only from the query terms in order to adapt search interfaces and to complement and enhance heavy processing tasks such as consulting specific search indexes.
  
 All groups of features contain valuable information about the input query as demonstrates the feature impact analysis. Features based on query projections on titles of Wikipedia articles, Freebase and Wikipedia category graph are the most relevant groups of features. However, the EIC performance can be improved if we include positive examples targeting other specific information repositories besides Wikipedia, such as IMDB or Last.fm.  This aspect will be taking into account in future research, as well as, the use of more complex methods for query pre-processing, such as spelling correctors and topic modeling.

We see this research as a first approach towards more sophisticated encyclopedic intent classifiers and specific applications, specially in direct search results.

\section{Acknowledgments}
This work was partially supported by Labs Sapo from Portugal Telecom. \\

\bibliographystyle{unsrt}
\bibliography{refs} 

\end{document}